\documentclass[12pt]{article}

\usepackage{epsfig,graphics,colordvi,color}
\usepackage{amsmath}\usepackage{amssymb}
\usepackage{hyperref}

\begin{document}

\title{Spin observables in pion photoproduction\\
from a unitary and causal effective field theory}
\author{A.M. Gasparyan$^{1,2}$, M.F.M. Lutz$^1$
\\ $^1$GSI Helmholtzzentrum f\"ur Schwerionenforschung GmbH,\\
Planckstrasse 1, 64291 Darmstadt, Germany
\\ $^2$SSC RF ITEP, Bolshaya Cheremushkinskaya 25,\\ 117218 Moscow,
š Russia}
\maketitle

\begin{abstract}
Pion photoproduction is analyzed with the chiral Lagrangian.
Partial-wave amplitudes are obtained by an analytic extrapolation of
subthreshold reaction amplitudes computed
in chiral perturbation theory, where the constraints set by
electromagnetic-gauge invariance, causality and
unitarity are used to stabilize the extrapolation. The experimental data
set is reproduced up to
energies $\sqrt{s}\simeq 1300$ MeV in terms of the parameters relevant
at order $Q^3$. We present and discuss predictions
for various spin observables.
\end{abstract}

\section{Introduction}

Chiral perturbation theory is a systematic tool for studying low-energy
hadron dynamics.
Particularly pion-nucleon scattering and pion photoproduction were
considered in
\cite{Bernard:1995dp,Bernard:2007zu,Bernard:1996gq,Fettes:1998ud}.
The application of $\chi $PT is however limited to the near threshold
region.
A method to extrapolate $\chi $PT results beyond the threshold region
using analyticity and unitarity constraints was proposed recently in
\cite{Gasparyan:2010xz}. We focus on results obtained for
pion photoproduction.
The predictions for spin observables as currently being measured at MAMI
are confronted with previous theoretical predictions.

\section{Chiral symmetry, causality and unitarity}

Our approach is based on the chiral Lagrangian involving pion,
nucleon and photon fields \cite{Fettes:1998ud,Bernard:2007zu}.
The terms relevant at the order $Q^3$ for pion elastic scattering and
pion photoproduction are listed
below\footnote{Note a typo in Eq.~(1) of \cite{Gasparyan:2010xz}. }
\begin{eqnarray}
\mathcal{L}_{int}&=&
-\frac{1}{4\,f^2}\,\bar{N}\,\gamma^{\mu}\,\big( \vec{\tau} \cdot
\big(\vec{\pi}\times
(\partial_\mu\vec{\pi})\big)\big) \,N +
\frac{g_A}{2\,f} \,\bar{N}\,\gamma_5\,\gamma^{\mu} \,\big(
\vec{\tau}\cdot (\partial_{\mu}\vec{\pi} )\big) \,N
\nonumber \\
&-&e\,\Big\{ \big(\vec{\pi}\times(\partial_{\mu}\vec{\pi}) \big)_3
+ \bar{N}\,\gamma_\mu\, \frac{1+\tau_3}{2} \,N
- \frac{g_A}{2\,f}
\,\bar{N}\,\gamma_5\,\gamma_{\mu}\,\big(\vec\tau\times\vec{\pi}\big)_3\,N \Big\}
\,A^\mu
\nonumber\\
&-&\frac{e}{4\,m_N}\,\bar{N}\,\sigma_{\mu\nu}\,\frac{\kappa_s+\kappa_v\,\tau_3}{2}\,N\,F^{\mu\nu}+
\frac{e^2}{32\pi^2
f}\,\epsilon^{\mu\nu\alpha\beta}\,\pi_3\,F_{\mu\nu}\,F_{\alpha\beta}
\nonumber\\
&-&\frac{2\,c_1}{f^2}\,m_\pi^2\, \bar{N}\,( \vec{\pi}\cdot\vec{\pi})\,N -
\frac{c_2}{2\,f^2\,m_N^2}\,\Big\{\bar{N}\,(
\partial_{\mu}\,\vec{\pi})\cdot (\partial_{\nu}\vec{\pi})\,(\partial^\mu
\partial^\nu N )+\rm{h.c.}\Big\}
\nonumber \\
&+& \frac{c_3}{f^2}\,\bar{N} \,(\partial_{\mu}\,\vec{\pi} )
\cdot (\partial^{\mu}\vec{\pi})\,N
-\frac{c_4}{2\,f^2}\,\bar{N}\,\sigma^{\mu\nu}\,\big(\vec{\tau} \cdot
\big((\partial_{\mu}\vec{\pi})\times
(\partial_{\nu}\vec{\pi})\big)\big)\,N
\nonumber\\
&-&i\,\frac{d_1+d_2}{f^2\,m_N}\,
\bar{N}\,\big(\vec\tau\cdot \big((\partial_\mu \vec \pi )\times
(\partial_\nu\partial_\mu \vec\pi) \big)\big) \, (\partial^\nu N) +
\rm{h.c.}
\nonumber \\
&+&\frac{i\,d_3}{f^2\,m_N^3}\,
\bar{N}\,\big(\vec \tau \cdot \big( (\partial_\mu\vec\pi )\times
(\partial_\nu \partial_\lambda\vec\pi )\big)\big)\,
(\partial^\nu\partial^\mu\partial^\lambda N)
+\mbox{h.c.}
\nonumber\\
&-&2\,i\,\frac{m_\pi^2\,d_5}{f^2\,m_N}\,\bar{N}\,\big(\vec{\tau} \cdot
\big(\vec{\pi}\times
(\partial_\mu\vec{\pi}) \big)\big)\,( \partial ^\mu N) +\rm{h.c.}
\nonumber\\
&-&\frac{i \,e }{f\,m_N }\, \epsilon^{\mu\nu\alpha\beta}\,\bar{N}\,\big(
d_8\,
(\partial_\alpha \,\pi_3) +d_9\,\big(\vec\tau \cdot (\partial_\alpha
\vec \pi)\big) \big)\, (\partial_\beta\,N)\, F_{\mu\nu}+\mbox{h.c.}
\nonumber \\
&+&i\,\frac{d_{14}-d_{15}}{2\,f^2\,m_N}\,
\bar{N}\,\sigma^{\mu\nu}\,\big((\partial_\nu\vec\pi )\cdot
(\partial_\mu\partial_\lambda\vec\pi ) \big)\,(\partial^\lambda N)
+\mbox{h.c.}
\nonumber \\
&-&\frac{m_\pi^2\,d_{18}}{f} \,\bar{N}\,\gamma_5\,\gamma^{\mu} \, \big(
\vec{\tau}\cdot (\partial_{\mu}\vec{\pi})\big) \, N
-\frac{e\,m_\pi^2\,d_{18}}{f} \,\bar{N}\,\gamma_5\,\gamma^{\mu} \, \big(
\vec{\tau}\times \vec{\pi}\big)_3 \, N\,A_\mu
\nonumber \\
&+&\frac{e \,(d_{22}-2\,d_{21})}{2\,f}\,\bar{N}\,\gamma_5\,\gamma^\mu\,
\big(\vec\tau\times\partial^\nu \,\vec \pi\big)_3 \,N\, F_{\mu\nu}
\nonumber \\
&+&\frac{e\, d_{20}}{2 \,f\,m_N^2}\,\bar{N}\,\gamma_5\,\gamma^\mu\,
\big(\vec\tau \times (\partial_\lambda \,\vec \pi)\big)_3\,
(\partial^\nu\partial^\lambda N)\, F_{\mu\nu}+\mbox{h.c.} \,.
\label{Lagrangian}
\end{eqnarray}
A strict chiral expansion of the amplitude to the order $Q^3$ includes
tree-level graphs, loop diagrams, and counter terms. Counter terms depend
on a few unknown parameters, which we adjusted to the empirical data on
$\pi N$
elastic scattering and pion photoproduction.
The extrapolation of the amplitudes obtained within ChPT is performed
utilizing constraints imposed by basic principles of analyticity and
unitarity.
For each partial wave we solved the non-linear integral equation
\begin{eqnarray}
&& T_{ab}^{JP}(\sqrt{s}\,)=U_{ab}^{JP}(\sqrt{s}\,)
\nonumber\\
&& \qquad +\sum_{c,d} \int_{\mu_{\rm
thr}}^{\infty}\frac{dw}{\pi}\frac{\sqrt{s}-\mu_M}{w-\mu_M}
\frac{T_{ac}^{*,JP}(w)\,\rho^{JP}_{cd}(w)\,T_{db}^{JP}(w)}{w-\sqrt{s}-i\epsilon}\,,
\label{disrel}
\end{eqnarray}
where the generalized potential, $U_{ab}^{(JP)}(\sqrt{s}\,)$, is the
part of the amplitude that contains left-hand cuts only. The phase-space
matrix $\rho^{JP}_{cd}(w)$ reflects our particular convention for the
partial-wave amplitudes, that are free of kinematical constraints. The
matching scale $\mu_M$ is required as to arrive at approximate crossing
symmetric results. For more details we refer to \cite{Gasparyan:2010xz}.

\section{Results}

\begin{figure}[t]
 \includegraphics*[width=13.5cm]{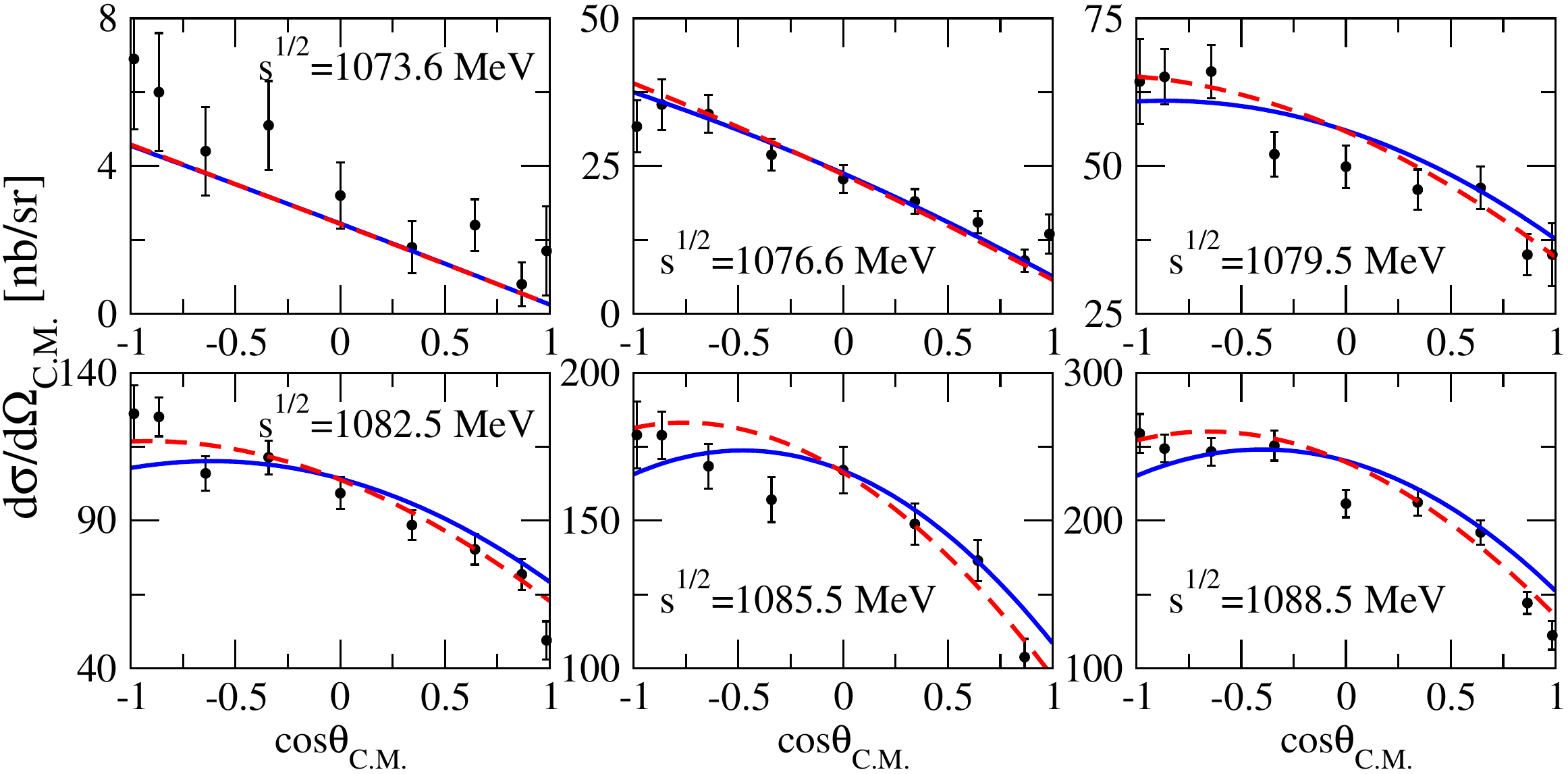}
\caption{Near threshold differential cross section for the
reaction $\gamma p\to\pi^0 p$ with data taken from
\cite{Schmidt:2001,Schmidt:2001vg}.
Shown are results from our coupled-channel theory including isospin
breaking effects as are implied by
the use of empirical pion
and nucleon masses. The solid lines correspond to our calculation with
only $s$- and $p$-wave multipoles
included. The effect of higher partial waves is shown by the dashed lines. }
\label{fig:pi0threshold}
\end{figure}

\begin{figure}[t]
 \includegraphics*[width=10.5cm]{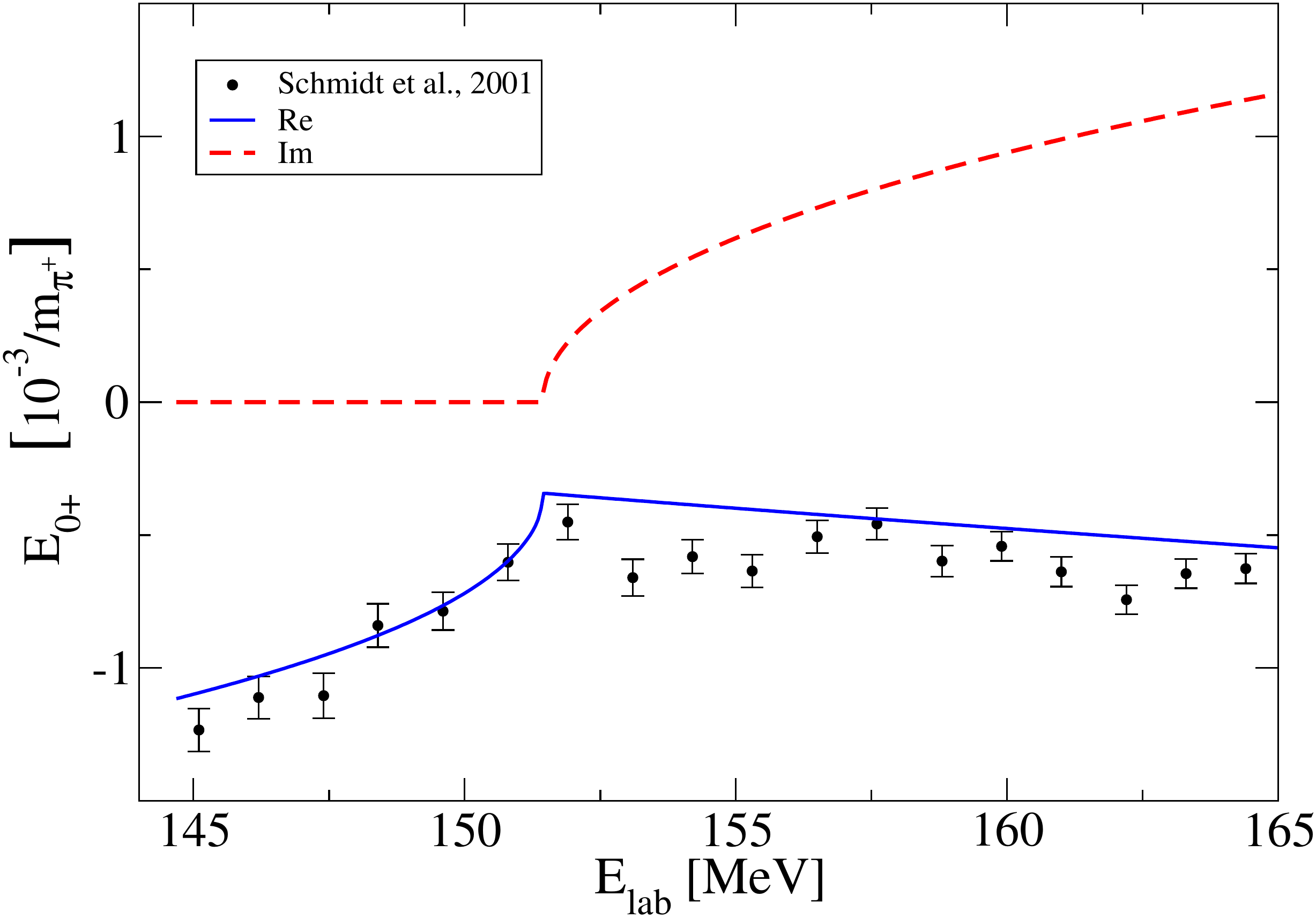}
\caption{Energy dependence of the $E_{0+}$ multipole close to the pion
production threshold. The data
are from \cite{Schmidt:2001vg}.}
\label{fig:E0p}
\end{figure}

The low-energy constants relevant for the elastic pion-nucleon
scattering were determined in \cite{Gasparyan:2010xz}.
The empirical $s$- and $p$-wave phase shifts are well reproduced up to
the energy $\sqrt{s}\approx 1300$ MeV. Above
this energy inelastic channels become important. The only exception is
the $P_{11}$ partial wave where the influence
of inelastic channels is significantly larger, that results in a
slightly worse description of the phase shift. A
convincing convergence pattern when going from $Q^1$ to $Q^3$
calculation was observed. In addition to the
low-energy constants of the Lagrangian (\ref{Lagrangian}) there are CDD
pole parameters characterizing the
Delta and Roper resonances \cite{Gasparyan:2010xz}.

\begin{figure}[t]
 \includegraphics*[width=13.5cm]{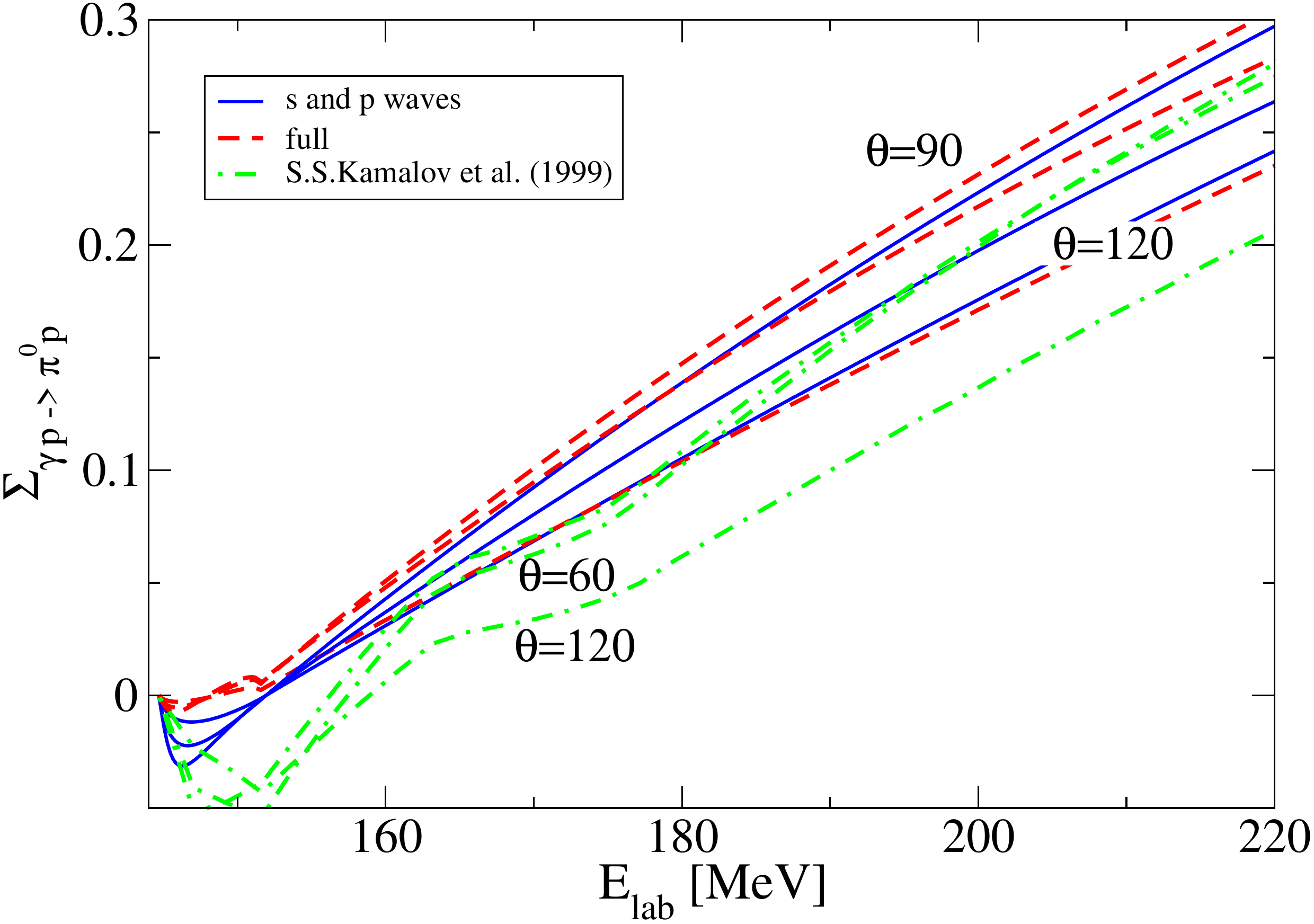}
\caption{Energy dependence of the beam asymmetry for three particular
angles.}
\label{fig:Sigma60_90_120}
\end{figure}

The pion-photoproduction $s$- and $p$-wave multipoles are quite
constrained. There are only four additional
low-energy constants and four CDD-pole parameters for the twelve
multipoles to be reproduced. Nevertheless a
good agreement to the existing partial wave analyzes was achieved. In
order to avoid the ambiguities in the different
partial-wave analyzes we determined the parameters from the experimental
data directly, where we excluded the
near-threshold data in the fit. Our results for the differential cross
sections, beam asymmetries, and helicity
asymmetry for the reaction channels $\gamma p\to\pi^0 p$, $\gamma
p\to\pi^+ n$, $\gamma n\to\pi^- p$ are in
agreement with experimental data from threshold up to $\sqrt{s}=1300$
MeV. Fig.~\ref{fig:pi0threshold} confronts
our prediction for the neutral pion-photo production with the
near-threshold MAMI data~\cite{Schmidt:2001,Schmidt:2001vg}.
This data allows one to extract the electric $s$-wave multipole
$E_{0+}$. Its energy dependence reveals a prominent
cusp effect at the opening of the $\pi^+ n$ channel. As shown
Fig.~\ref{fig:E0p} this structure is reproduced
by our calculation, which discriminates the channels with neutral and
charged pions.

\begin{figure}[t]
 \includegraphics*[width=12.8cm]{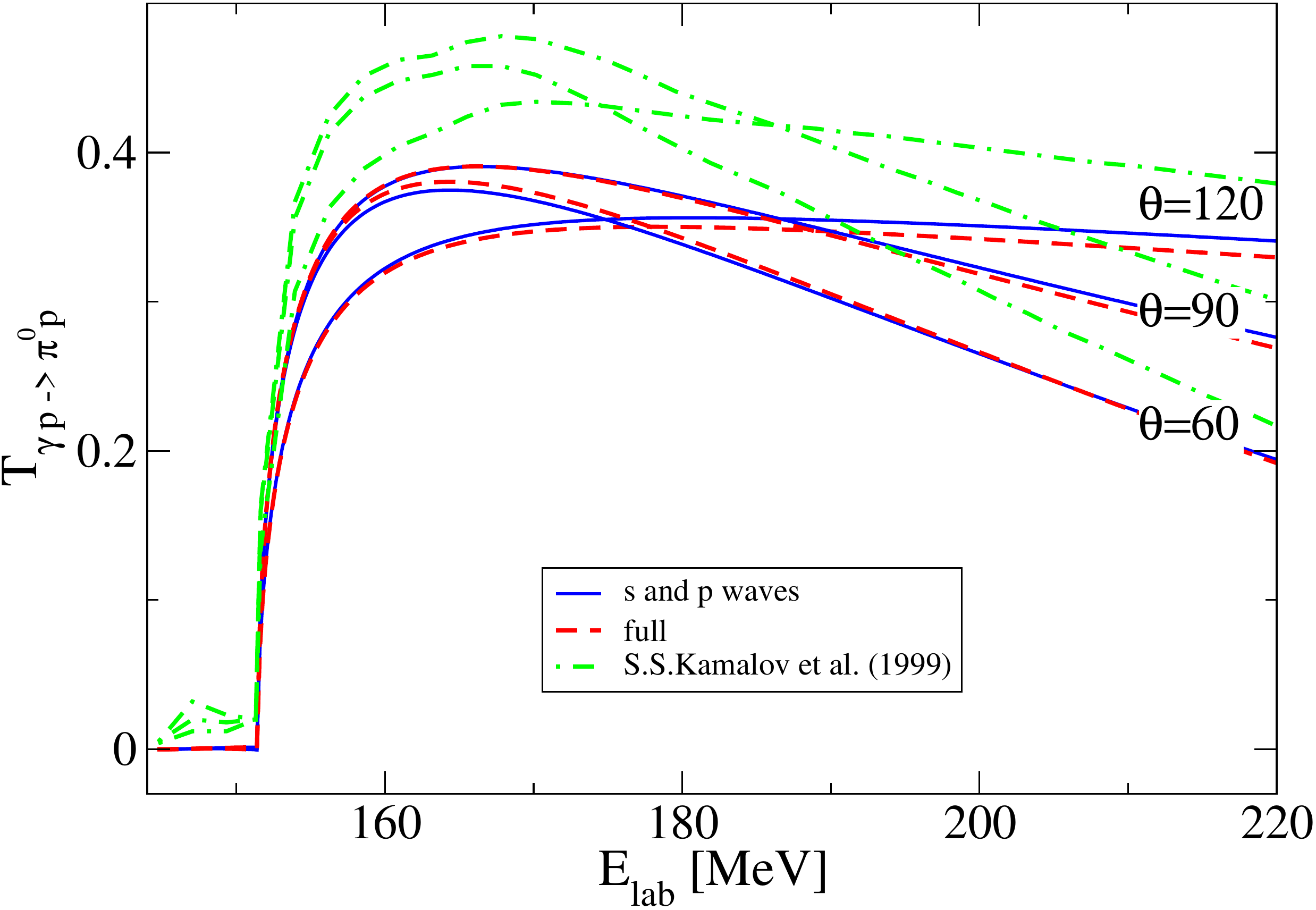}
\caption{Energy dependence of the target asymmetry
for three particular angles.}
\label{fig:T60_90_120}
\end{figure}

\begin{figure}[t]
 \includegraphics*[width=12.5cm]{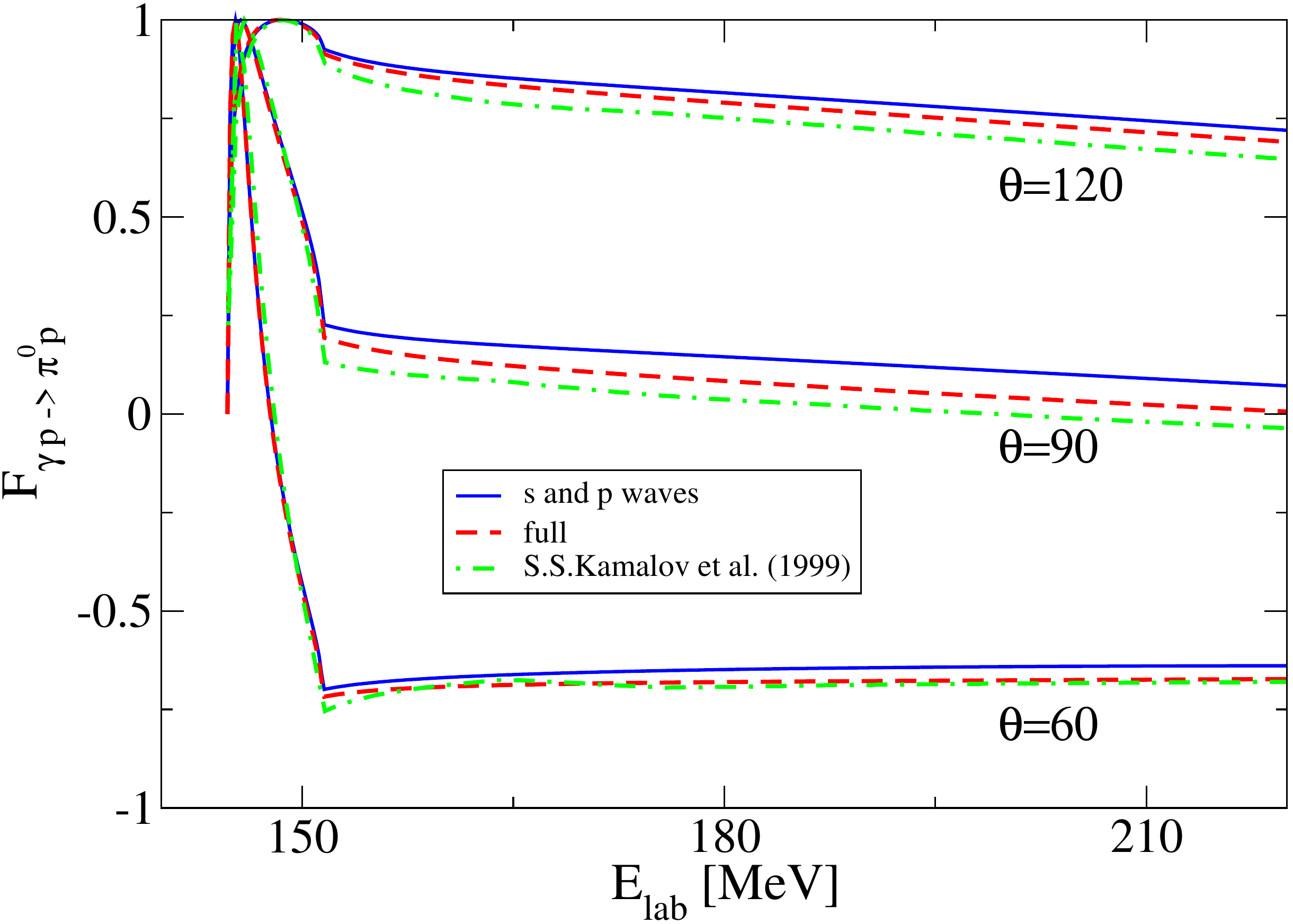}
\caption{Energy dependence of the double
polarization observable $F$
for three particular angles.}
\label{fig:F60_90_120}
\end{figure}

Further information about the low-energy photoproduction dynamics
is encoded in the $p$-wave threshold multipoles. In order to disentangle
the three independent $p$-wave amplitudes it is insufficient to measure
the differential cross section only. The near-threshold beam asymmetry in $\gamma
p\to\pi^0 p$ was measured by MAMI \cite{Schmidt:2001vg}. Our results are in striking disagreement
with that measurement.
We predict the beam asymmetry to change sign close to threshold, in a
similar manner as predicted before in the dynamical model of Kamalov et al.
\cite{Kamalov:2001qg}. In Fig. \ref{fig:Sigma60_90_120} we compare the
two different results on the beam
asymmetry. Though there is qualitative agreement, important quantitative
differences remain. It is interesting
to observe that d-wave multipoles appear to play an important role in
the near-threshold region.
This was discussed also in \cite{FernandezRamirez:2009jb}. Thus the beam
asymmetry may not be the optimal quantity to extract the p-wave
threshold amplitudes. Currently a new data set on the beam asymmetry is being analyzed at MAMI.

We conclude that it is important to take further data on spin observables
other than the beam asymmetry.
Two cases are currently been studied at MAMI close to threshold. The
target asymmetry $T$ and the
double polarization observable $F$. Since there are different phase
conventions used in the literature the
reader may appreciate that we detail the relevant expression in the
convention used in
\cite{Gasparyan:2010xz}. It holds
\begin{eqnarray}
&& \frac{d\sigma}{d\Omega}=\frac{\bar{p}_{\rm cm}}{2\,p_{\rm cm}}
\left(|H_N|^2+|H_{SA}|^2+|H_{SP}|^2+|H_D|^2\right)\,,
\nonumber\\
&& \Sigma\,\frac{d\sigma}{d\Omega}=\frac{\bar{p}_{\rm cm}}{p_{\rm cm}}\,
\Re\left(H_{SP}H_{SA}^*-H_NH_D^*\right)\,,
\nonumber\\
&& T\,\frac{d\sigma}{d\Omega}=\frac{\bar{p}_{\rm cm}}{p_{\rm cm}}\,
\Im\left(H_{SP}H_{N}^*+H_DH_{SA}^*\right)\,,
\nonumber\\
&&F\,\frac{d\sigma}{d\Omega}=\frac{\bar{p}_{\rm cm}}{p_{\rm cm}}\,
\Re\left(H_{SP}H_{N}^*+H_DH_{SA}^*\right)\,,
\end{eqnarray}
with $x= \cos \theta$ and
\begin{eqnarray}
&& H_{N\;\;}=\frac{m_N}{4\pi\,\sqrt{s}}\,\cos \frac{\theta}{2}\,
\sum_J\left(t^J_{+,1}-t^J_{-,1}\right)
\left(P'_{J+\frac{1}{2}}(x) -P'_{J-\frac{1}{2}}(x)\right) \,,
\nonumber\\
&& H_{SA}=-\frac{m_N}{4\pi\,\sqrt{s}}\,\sin \frac{\theta}{2}\,
\sum_J \left(t^J_{+,1}+t^J_{-,1}\right)
\left(P'_{J+\frac{1}{2}}(x)+P'_{J-\frac{1}{2}}(x)\right) \,,
\nonumber\\
&& H_{SP}=\frac{m_N}{4\pi\,\sqrt{s}}\,\frac{-\sin\theta
\cos{\frac{\theta}2}}{\sqrt{(J-\frac{1}{2})(J+\frac{3}{2})}}\,
\sum_J \left(t^J_{+,2}-t^J_{-,2}\right)
\left(P''_{J+\frac{1}{2}}(x) -P''_{J-\frac{1}{2}}(x)\right) \,,
\nonumber\\
&& H_{D\;\;}=\frac{m_N}{4\pi\,\sqrt{s}}\,\frac{\sin\theta \sin
{\frac{\theta}2}}{\sqrt{(J-\frac{1}{2})(J+\frac{3}{2})}}\,
\sum_J \left(t^J_{+,2}+t^J_{-,2}\right)
\left(P''_{J+\frac{1}{2}}(x)
+P''_{J-\frac{1}{2}}(x)\right)\,. \nonumber
\label{}
\end{eqnarray}

In Fig.~\ref{fig:T60_90_120} an Fig.~\ref{fig:F60_90_120} we
show our predictions for the energy dependence of the target
asymmetry and the $F$ observable in comparison with the predictions of the
dynamical model of Ref.~\cite{Kamalov:2001qg} calculated at different
angles. One can see that the target and beam asymmetry are most sensitive to the details of
the dynamics, whereas the $F$ observable is quite similar in both approaches.

In order to unravel the dynamics close to threshold we detail the contributions of
$s$- and $p$-waves to the differential cross section and the  $\Sigma$, $T$ and $F$ observables. It holds
\begin{eqnarray}
\frac{d\sigma}{d\Omega}&=&\frac{\bar{p}_{\rm cm}}{p_{\rm cm}}\,\Big[|E_{0+}|^2+\frac{1}{2}\,|P_2|^2+\frac{1}{2}\,|P_3|^2
+2\,\Re(E_{0+}\,P_1^*)\,\cos{\theta}
\nonumber\\
&+&\Big(|P_1|^2-\frac{1}{2}\,|P_2|^2-\frac{1}{2}\,|P_3|^2\Big)\,\cos^2{\theta}\Big]\, ,
\nonumber\\
\Sigma\,\frac{d\sigma}{d\Omega}&=&\frac{\bar{p}_{\rm cm}}{2\,p_{\rm cm}}\,\big[|P_3|^2-|P_2|^2\big]\,\sin^2{\theta}\, ,
\nonumber\\
T\,\frac{d\sigma}{d\Omega}&=&-\frac{\bar{p}_{\rm cm}}{p_{\rm cm}}\,
\Im{\big[ ( E_{0+}+P_1 \cos{\theta})\,(P_2-P_3)^*\big] }\sin{\theta}\, ,
\nonumber\\
F\,\frac{d\sigma}{d\Omega}&=&\frac{\bar{p}_{\rm cm}}{p_{\rm cm}}\,
\Re{\big[ ( E_{0+}+P_1 \cos{\theta})\,(P_2-P_3)^*\big] }\sin{\theta},
\label{dwaves}
\end{eqnarray}
with the linear combinations of $p$-wave multipoles $P_1=3\,E_{1+}+M_{1+}-M_{1-}$,
$P_2=3\,E_{1+}-M_{1+}+M_{1-}$, $P_3=2\,M_{1+}+M_{1-}$.

The expression for the target asymmetry $T$ (Eq.~(\ref{dwaves}))
depends on the imaginary parts of the multipoles. That is why close
to the $\pi^0 p$ threshold $T$ is very small (see Fig.~\ref{fig:T60_90_120}). Slightly above the
$\pi^+ n$ threshold it holds approximately
\begin{eqnarray}
T\,\frac{d\sigma}{d\Omega}(\theta=90^\circ)\approx
-\frac{\bar{p}_{\rm cm}}{p_{\rm cm}}\,
\Im{( E_{0+})}\,(P_2-P_3)\,,
\end{eqnarray}
since the imaginary parts of $p$-wave multipoles are small.
The imaginary part of $E_{0+}$ is in turn dominated by the intermediate $\pi^+ n$ state.
This allows one to access the difference $P_2-P_3$ in the vicinity of the $\pi^+ n$ threshold.

In the beam asymmetry $\Sigma$ the  $p$-waves multipoles enter only quadratically and therefore terms containing an
interference of the $E_{0+}$ and $d$-wave amplitudes are not suppressed by powers of
the $\pi N$ momentum. Moreover the magnitudes of $|P_2|$ and $|P_3|$ are similar \cite{Schmidt:2001vg} which
further diminishes the relative importance of the $p$-wave contribution to $\Sigma$. In contrast, the quantity
$F(\theta=90^\circ)$ contains the term $\Re{\left(E_{0+}\,(P_2 - P_3)^*\right)}$ but no other competing
contribution. Thus measuring $F$ provides a reliable determination
of $P_2-P_3$. Note that this difference is not small because $P_2$ and $P_3$ have opposite signs
\cite{Schmidt:2001vg}.

\section{Summary}

We studied pion photoproduction from threshold up to $\sqrt{s}=1300$
MeV with a novel approach developed in  \cite{Gasparyan:2010xz} based on an analytic extrapolation
of subthreshold amplitudes calculated in ChPT. The free parameters were adjusted to the pion-nucleon
and photoproduction empirical data excluding the threshold region. Nevertheless the near-threshold
MAMI data on the reaction $\gamma p\to\pi^0 p$ are described well. The energy dependence of the
$s$-wave electric $E_{0+}$ multipole close to threshold including its prominent
cusp structure is also well reproduced. We presented predictions for spin observables that are
planned to be measured or being analyzed at MAMI. Our predictions are compared with results of the dynamical
model of Kamalov et al. \cite{Kamalov:2001qg}. The importance of $d$-waves for the beam asymmetry close to
the $\pi N$ threshold was emphasized. This effect indicates that the beam asymmetry may be not
best suited to disentangle the various $p$-wave threshold multipoles as has been anticipated
before. We argued that the measurement of the double polarization observable $F$ suits this
purpose much better.

\vskip1cm
{\bfseries{Acknowledgments}}
\vskip0.4cm
We thank M. Ostrick and L. Tiator for stimulating discussions.

\clearpage

 \bibliographystyle{elsart-num}
\bibliography{1}

\begin{thebibliography}{1}
\expandafter\ifx\csname url\endcsname\relax
  \def\url#1{\texttt{#1}}\fi
\expandafter\ifx\csname urlprefix\endcsname\relax\def\urlprefix{URL }\fi

\bibitem{Bernard:1995dp}
V.~Bernard, N.~Kaiser, U.-G. Meissner, {Chiral dynamics in nucleons and
  nuclei}, Int. J. Mod. Phys. E4 (1995) 193--346.

\bibitem{Bernard:2007zu}
V.~Bernard, {Chiral Perturbation Theory and Baryon Properties}, Prog. Part.
  Nucl. Phys. 60 (2008) 82--160.

\bibitem{Bernard:1996gq}
V.~Bernard, N.~Kaiser, U.-G. Meissner, {Determination of the low-energy
  constants of the next-to- leading order chiral pion nucleon Lagrangian},
  Nucl. Phys. A615 (1997) 483--500.

\bibitem{Fettes:1998ud}
N.~Fettes, U.-G. Meissner, S.~Steininger, {Pion nucleon scattering in chiral
  perturbation theory. I: Isospin-symmetric case}, Nucl. Phys. A640 (1998)
  199--234.

\bibitem{Gasparyan:2010xz}
A.~Gasparyan, M.~Lutz, {Photon- and pion-nucleon interactions in a unitary and
  causal effective field theory based on the chiral Lagrangian}, Nucl.Phys.
  A848 (2010) 126--182.

\bibitem{Schmidt:2001}
A.~Schmidt, PhD thesis, Mainz (2001), http://wwwa2.kph.uni-maiz.de/A2/.

\bibitem{Schmidt:2001vg}
A.~Schmidt, et~al., {Test of low-energy theorems for p(gamma(pol.),pi0)p in the
  threshold region}, Phys. Rev. Lett. 87 (2001) 232501.

\bibitem{Kamalov:2001qg}
S.~S. Kamalov, G.-Y. Chen, S.-N. Yang, D.~Drechsel, L.~Tiator, {pi0 photo- and
  electroproduction at threshold within a dynamical model}, Phys. Lett. B522
  (2001) 27--36.

\bibitem{FernandezRamirez:2009jb}
C.~Fernandez-Ramirez, A.~M. Bernstein, T.~W. Donnelly, {The unexpected impact
  of D waves in low-energy neutral pion photoproduction from the proton and the
  extraction of multipoles}, Phys. Rev. C80 (2009) 065201.

\end{thebibliography}

\end{document}